# Optimal Non-blocking Decentralized Supervisory Control Using G-Control Consistency

Vahid Saeidi[a], Ali A. Afzalian[*b], Davood Gharavian[c]

[*] Phone +982173932626, Fax +982177310425
[a,b,c] Department of Electrical Eng., Abbaspour School of Engineering,
Shahid Beheshti University, Tehran, Iran
[a] v_saeidi@sbu.ac.ir, [b] Afzalian@sbu.ac.ir, [c] d_gharavian@sbu.ac.ir

**Abstract**

Supervisory control synthesis encounters with computational complexity. This can be reduced by decentralized supervisory control approach. In this paper, we define intrinsic control consistency for a pair of states of the plant. G-control consistency (GCC) is another concept which is defined for a natural projection w.r.t. the plant. We prove that, if a natural projection is output control consistent for the closed language of the plant, and is a natural observer for the marked language of the plant, then it is G-control consistent. Namely, we relax the conditions for synthesis the optimal non-blocking decentralized supervisory control by substituting GCC property for L-OCC and $L_m$-observer properties of a natural projection. We propose a method to synthesize the optimal non-blocking decentralized supervisory control based on GCC property for a natural projection. In fact, we change the approach from language-based properties of a natural projection to DES-based property by defining GCC property.

## 1. Introduction[1]

Supervisory control synthesis for a monolithic specification encounters with computational complexity, in discrete-event systems (DES).

Modular (Willner & Heymann, 1991; Malik & Teixeira, 2016; Komenda, Masopust & van Schuppen, 2012), hierarchical (Zhong & Wonham, 1990), heterarchical (Schmidt, Moor & Perk, 2008 ; Schmidt & Breindl, 2011), synthesis methods and non-deterministic automata approaches (Malik, Flordal & Pena, 2007; Mohajerani, Malik & Fabian, 2014; Mohajerani, Malik & Fabian, 2017), have been proposed, in order to tackle the computational complexity.

Decentralized supervisory control has been introduced to reduce the computational complexity in large scale DES (Rudie & Wonham, 1992; Yoo & Lafortune, 2002; Lin & Wonham, 1990; Cai, Zhang & Wonham, 2015). Since, a decentralized supervisor partially observes the plant; it does not have enough information about actions of other supervisors. Hence, there may be conflict between decentralized supervisors. Conditions were found in (Lin & Wonham, 1988), for equivalency between decentralized and monolithic supervisory control, where prefix closed languages are only considered and the non-blocking problem are not addressed. Since normality must be verified for the monolithic supervisor, their conditions are not easy to check. To solve this problem, a method was introduced in (Feng & Wonham, 2008), to synthesize the optimal (least restrictive) non-blocking decentralized supervisory control based on the observer property for a natural projection of the marked language of the plant, and the output control consistency (OCC) property for the closed language of the plant. A local version of OCC (i.e. LCC) was defined in (Schmidt & Breindl, 2011), to be employed in the maximal permissive hierarchical supervisory control.

Decomposability and strong decomposability (conormality) were introduced in (Rudie & Wonham, 1992), to construct decentralized supervisory control in a top-down approach.

In the bottom-up approach, construction of a coordinator was proposed to remove the conflict between decentralized supervisors (Feng & Wonham, 2008; Komenda, Ma-sopust & van Schuppen, 2015; Wonham, 2016). Also, some research works have been reported on detecting conflict between decentralized supervisors using the observer property of a natural projection (Pena, Cury & Lafortune, 2009).

---

[1] This paper was submitted to the **Automatica** in February 16, 2017. Currently, it is under review for possible publication.

OCC is a strong property, which was proposed for synthesis the optimal non-blocking decentralized supervisory control, in the literature (Feng & Wonham, 2008). This property is so strong that can rarely be satisfied in practice. That is because OCC is a property of a language. Here, we are going to define a more relaxed property instead of OCC.

In this paper, we propose a method to synthesize the optimal non-blocking decentralized supervisory control based on projection of a reduced model of the plant, instead of direct projection of the closed and the marked languages of the plant, proposed in (Feng & Wonham, 2008). By this method, we have a decentralized supervisory controller, which is control equivalent to a monolithic supervisor using the same specification.

It is proved that the projection of a DES can be employed to synthesize the optimal non-blocking decentralized supervisory control, provided that the natural projection is G-control consistent (GCC).

In order to define GCC property for a natural projection, it is necessary to introduce the intrinsic control consistency (ICC) of a pair of states. Each pair of states in a DES is ICC provided that (i) any controllable event which is defined at one state, is not defined at the other, (ii) they are not both marked.

A natural projection is GCC, provided that each pair of states in the plant, which are reachable by lookalike strings, are ICC.

We prove that if a natural projection is both L-OCC and $L_m$-observer for the plant, then it is GCC.

The rest of the paper is organized as follows: In Section 2, the necessary preliminaries and basic notions of supervisory control theory are reviewed. In Section 3, a new observation property (i.e. G- control consistency) for a natural projection is defined. In Section 4, a method is proposed to synthesize the optimal non-blocking decentralized supervisory control based on GCC property for the natural projection. In Section 5, the proposed method is employed to synthesize the optimal non-blocking decentralized supervisory control for surge-avoidance in a gas compressor station. Finally, concluding remarks are outlined in Section 6.

## 2. Preliminaries

A discrete-event system is presented by an automaton $\mathbf{G} = (Q, \Sigma, \delta, q_0, Q_m)$, where $Q$ is a finite set of states, with $q_0 \in Q$ as the initial state and $Q_m \subseteq Q$ being the marked states; $\Sigma$ is a finite set of events ($\sigma$), which is partitioned as a set of controllable events $\Sigma_c$ and a set of uncontrollable events $\Sigma_u$, where $\Sigma = \Sigma_c \cup \Sigma_u$. $\Sigma^+$ denotes the set of all finite strings of events. Also, $\Sigma^* := \{\epsilon\} \cup \Sigma^+$, where $\epsilon$ is the empty string. $\delta$ is a transition mapping $\delta: Q \times \Sigma \to Q$, $\delta(q, \sigma) = q'$ gives the next state $q'$ is reached from $q$ by the occurrence of $\sigma$. In this context $\delta(q_0, s)!$ means that $\delta$ is defined for $s$ at $q_0$. $L(\mathbf{G}) := \{s \in \Sigma^* | \delta(q_0, s)!\}$ is the closed behavior of $\mathbf{G}$ and $L_m(\mathbf{G}) := \{s \in L(\mathbf{G}) | \delta(q_0, s) \in Q_m\}$ is the marked behavior of $\mathbf{G}$ (Wonham, 2016). $\Sigma^*$ is the set of all strings of events from $\Sigma$, and the symbol "$s$" represents one such string.

In the supervisory control context, all events $\Sigma$ are partitioned as a set of controllable events $\Sigma_c$ and a set of uncontrollable events $\Sigma_u$, where $\Sigma = \Sigma_c \cup \Sigma_u$. A control pattern is $\gamma$, where $\Sigma_u \subseteq \gamma \subseteq \Sigma$ and the set of all control patterns is denoted with $\Gamma = \{\gamma \in 2^\Sigma | \gamma \supseteq \Sigma_u\}$. A supervisor for a generator $\mathbf{G}$ is a map $V: L(\mathbf{G}) \to \Gamma$, where $V(s)$ represents the set of enabled events after the occurrence of the string $s \in L(\mathbf{G})$. A pair $(\mathbf{G}, V)$ is written as $V/\mathbf{G}$ and is called "$\mathbf{G}$ is under supervision by $V$". The closed loop language $L(V/\mathbf{G})$ is defined by: (1) $\epsilon \in L(V/\mathbf{G})$, (2) $s\sigma \in L(V/\mathbf{G})$ iff $s \in L(V/\mathbf{G}), \sigma \in V(s),$ and $s\sigma \in L(\mathbf{G})$. The marked language of $V/\mathbf{G}$ is $L_m(V/\mathbf{G}) = L(V/\mathbf{G}) \cap L_m(\mathbf{G})$. The closed loop system is non-blocking if $\overline{L_m(V/\mathbf{G})} = L(V/\mathbf{G})$. $\overline{L_m(V/\mathbf{G})}$ is the set of all prefixes of traces in $L_m(V/\mathbf{G})$. A language $K \subseteq \Sigma^*$ is controllable w.r.t. $L(\mathbf{G})$ and uncontrollable events $\Sigma_u$, if $\overline{K}\Sigma_u \cap L(\mathbf{G}) \subseteq \overline{K}$. The set of all controllable sublanguages $E$ w.r.t. $L(\mathbf{G})$ and $\Sigma_u$ is denoted by $C(E) = \{K \subseteq E | \overline{K}\Sigma_u \cap L(\mathbf{G}) \subseteq \overline{K}\}$, that is nonempty and closed under union. For every specification language $E$, there exists the supremal controllable sublanguage of $E$ w.r.t. $L(\mathbf{G})$ and $\Sigma_u$ (Wonham, 2016; Cassandras & Lafortune, 2008). Assume that $\mathbf{SUP} = (X, \Sigma, \xi, x_0, X_m)$ is the recognizer of $K \neq \emptyset$. If $\mathbf{G}$ and $\mathbf{E}$ are finite-state DES, then $K$ is regular language. Write $|\cdot|$ for the state size of DES. Then $|\mathbf{SUP}| \leq |\mathbf{G}||\mathbf{E}|$. In practice, engineers want to employ a reduced state supervisor, which has a fewer number of states, and is control equivalent to $\mathbf{SUP}$ w.r.t. $\mathbf{G}$ (Su & Wonham, 2004). A controller $\mathbf{RSUP}$ (which may be a reduced supervisor) is control equivalent to $\mathbf{SUP}$ w.r.t. $\mathbf{G}$, if (1) and (2) are satisfied.

$$L_m(\mathbf{G}) \cap L_m(\mathbf{RSUP}) = L_m(\mathbf{SUP}), \qquad (1)$$
$$L(\mathbf{G}) \cap L(\mathbf{RSUP}) = L(\mathbf{SUP}). \qquad (2)$$

The natural projection is a mapping $P: \Sigma^* \to \Sigma_0^*$, where (1) $P(\epsilon) := \epsilon$, (2) for $s \in \Sigma^*, \sigma \in \Sigma, P(s\sigma) := P(s)P(\sigma)$, and (3) $P(\sigma) := \sigma$ if $\sigma \in \Sigma_0$ and $P(\sigma) := \epsilon$ if $\sigma \notin \Sigma_0$. The effect of an arbitrary natural projection $P$ on a string $s \in \Sigma^*$ is to erase the events in $s$ that do not belong to observable events set, $\Sigma_0$. The natural projection $P$ can be extended and denoted by $P: Pwr(\Sigma^*) \to Pwr(\Sigma_0^*)$. For any $L \subseteq \Sigma^*$, $P(L) := \{P(s)|s \in L\}$. The inverse image function of $P$ is denoted by $P^{-1}: Pwr(\Sigma_0^*) \to Pwr(\Sigma^*)$ for any $L \subseteq \Sigma_0^*$, $P^{-1}(L) := \{s \in \Sigma^* | P(s) \in L\}$ (Wonham, 2016). The synchronous product of languages $L_1 \subseteq \Sigma_1^*$ and $L_2 \subseteq \Sigma_2^*$ is defined by $L_1 \| L_2 = P_1^{-1}(L_1) \cap P_2^{-1}(L_2) \subseteq \Sigma^*$, where $P_i: \Sigma^* \to \Sigma_i^*$, $i = 1,2$ for the union $\Sigma = \Sigma_1 \cup \Sigma_2$. $L_1$ and $L_2$ are synchronously non-conflicting, if $\overline{L_1 \| L_2} = \overline{L_1} \| \overline{L_2}$ (Feng & Wonham, 2008). If $L_1 = L_m(\mathbf{G_1})$ and $L_2 = L_m(\mathbf{G_2})$, then the procedure **sync** in TCT software (Wonham, 2014), returns $\mathbf{G} = \mathbf{sync}(\mathbf{G_1}, \mathbf{G_2})$, where $L_m(\mathbf{G}) = L_m(\mathbf{G_1}) \| L_m(\mathbf{G_2})$ and $L(\mathbf{G}) = L(\mathbf{G_1}) \| L(\mathbf{G_2})$ (Wonham, 2016).

Partial observation of events may remove critical information and cause inconsistent decision making by the supervisor. For instance, a non-blocking supervisor, synthesized based on full observation of the plant, could result a blocking supervisor under partial observation. To avoid this problem, the observable events of a plant must be carefully selected. In general, a subset of observable events does not need to have particular relation to a subset of controllable events.

A natural projection which has a "good" observable event set is called a natural observer (Feng & Wonham, 2008). Assume $L \subseteq \Sigma^*$ is the plant language and $\Sigma_0 \subseteq \Sigma$ is an observable event subset. The natural projection $P: \Sigma^* \to \Sigma_0^*$ is an $L$-observer if
$$(\forall s \in \bar{L})(\forall t \in P(L))P(s) \leq t \Rightarrow (\exists u \in \Sigma^*) su \in L \wedge P(su) = t$$

The symbol $\leq$ means that the first string is a prefix of the second.

In decentralized supervisory control, a decentralized controller observes and disables only events in an observable event subset. It was proved that the observer property is a sufficient condition for non-blocking decentralized control (Feng & Wonham, 2008). The authors extended this to achieve optimality (i.e. maximal permissiveness) using output control consistency (OCC) introduced in (Wong, Wonham, 1996), which has been used to study the hierarchical supervisory control with a general causal reporter map. The natural projection $P: \Sigma^* \to \Sigma_0^*$ is OCC for the prefix-closed language $L \subseteq \Sigma^*$, if for every string $s \in L$ of the form $s = \sigma_1 \sigma_2 \dots \sigma_k$ or $s = s' \sigma_1 \sigma_2 \dots \sigma_k$ $k \geq 1$ which satisfies the conditions that $s'$ terminates with an event in $\Sigma_0, \sigma_i \in \Sigma - \Sigma_0$ $(i \leq k-1)$ and $\sigma_k \in \Sigma_u \cap \Sigma_0$, then $(i \leq k)$ $\sigma_i \in \Sigma_u$. Here, the symbol $\leq$ means that the first variable is less than or equal to the second. Informally, when $\sigma_k$ is observable and uncontrollable, it's nearest controllable event must be observable.

The language $K$ is $(L(\mathbf{G}), P)$-normal, if $P^{-1}P(\bar{K}) \cap L(\mathbf{G}) = \bar{K}$. Paranormality is another property of $K$ w.r.t. $(L(\mathbf{G}), P)$. $K \subseteq \Sigma^*$ is $(L(\mathbf{G}), P) - paranormal$ if $\bar{K}(\Sigma - \Sigma_0) \cap L(\mathbf{G}) \subseteq \bar{K}$ (Wonham, 2016). It means that, $K$ is $(L(\mathbf{G}), P)$-paranormal, if the occurrence of unobservable events never exits the closure of $K$. If $\bar{K}$ is $(L(\mathbf{G}), P)$-normal, then it is $(L(\mathbf{G}), P) - paranormal$, too. But the reverse is not true. By analogy with controllability, the class of $(L(\mathbf{G}), P) - paranormal$ sublanguages of an arbitrary sublanguage of $\Sigma^*$ is nonempty, closed under union (but is not closed necessarily under intersection), and contains a (unique) supremal element (Wonham, 2016).

## 3. Observation Properties of a Natural Projection for DES model of a Plant

A natural projection $P$ can be imposed on a language $L$, which can be the closed language of a plant $\mathbf{G}$, (i.e. $L \coloneqq L(\mathbf{G})$). The effect of $P$ on the closed language ($L$), the marked language ($L_m$), and the DES model of the plant $\mathbf{G}$ can be described by $L(\mathbf{PG}) = PL(\mathbf{G})$ and $L_m(\mathbf{PG}) = PL_m(\mathbf{G})$ (Wonham, 2016).

In order to guarantee that a decentralized supervisory controller with partial observation is not blocking in the plant, we should find circumstances for natural projection, such that any synthesized decentralized controller makes consistent decisions at all states of the plant. We define intrinsic control consistency for each pair of states in the plant, by which any decentralized controller makes consistent decisions at each state of the plant. Arbitrary pair of states in the plant are intrinsic control consistent, if each disabled controllable event at one state is not enabled at the other, and they are not both marked. This concept is capsulated in Definition 1.

*Definition 1 (Intrinsic Control Consistent states):* Let $\mathbf{G} = (Q, \Sigma, \delta, q_0, Q_m)$ be a non-blocking plant. $q_i, q_j$ are intrinsic control consistent (ICC), if
$(i)$ $(\forall \sigma \in \Sigma_c)(\exists s \neq s' \in \Sigma^*), q_i = \delta(q_0, s),$
$\quad\quad q_j = \delta(q_0, s'),$
$\delta(q_i, \sigma)! \Rightarrow \neg \delta(q_j, \sigma)!, \delta(q_j, \sigma)! \Rightarrow \neg \delta(q_i, \sigma)!$ (2)
$(ii)$ $q_i \in Q_m \Rightarrow q_j \notin Q_m, q_j \in Q_m \Rightarrow q_i \notin Q_m$

Now, we introduce a definition to describe control properties of **PG** according to control properties of **G**. Let
$\Sigma_0 \coloneqq \Sigma$,
$\mathbf{G_0} \coloneqq \mathbf{G}$,
$\mathbf{G_k} \coloneqq \mathbf{P_k G_{k-1}}$, where $P_k: \Sigma_k^* \to \Sigma_{k+1}^*$, $\Sigma_{k+1} = \Sigma_k - \{\sigma_k\}$ and $\sigma_k \in \Sigma_u, k = 1, \dots, N$
$\mathbf{G_k} \coloneqq (Q_k, \Sigma_k, \delta_k, q_{0,k}, Q_{m,k})$,
The number of events belong to $\Sigma_n \coloneqq \bigcup_k \sigma_k$ is $N$.

We define a property, called $\mathbf{G_{k-1}} - Patially\ Control\ Consistent$ for a natural projection $P_k$.

*Definition 2 ($\mathbf{G_{k-1}} - Partially\ Control\ Consistent$ Projection):* A natural projection $P_k: \Sigma_k^* \to \Sigma_{k+1}^*$, is $\mathbf{G_{k-1}} - Partially\ Control\ Consistent$, if $(\forall i \in I^k, I^k$ is the index set$), q_i \in Q_{k-1}$ and $[\delta(q_i, \sigma_k^*)! \Rightarrow (\exists j), q_j = \delta(q_i, \sigma_k^*)]$ are true, then $\forall j, q_j \in Q_{k-1}$ are pairwise intrinsic control consistent states. Also, $\Sigma_{k+1} = \Sigma_k$, if $q_i, q_j$ are not intrinsic control consistent, for $\exists i, j \in I^k, i \neq j, q_i, q_j \in Q_{k-1}$ such that $q_j = \delta(q_i, \sigma_k^*)$.

In this case, we may find a set of all events $\sigma_k$, (i.e. $\Sigma_n$) such that their corresponding $P_k$ is $\mathbf{G_{k-1}} - Partially\ Control\ Consistent$.

*Definition 3 ($\mathbf{G} - Control\ Consistent\ Projection$):* A natural projection $P: \Sigma^* \to \Sigma_0^*$, with $\Sigma_c \subseteq \Sigma_0$ is $\mathbf{G} - Control\ Consistent$ (GCC), if $(\forall s, s' \in \Sigma^*)(\forall i, j \in I, I$ is the index set$)$ $q_i = \delta(q_0, s), q_j = \delta(q_0, s')$ and $P(s) = P(s')$, then $q_i, q_j$ are intrinsic control consistent.

*Proposition 1:* Let $\mathbf{G} = (Q, \Sigma, \delta, q_0, Q_m)$ and $P: \Sigma^* \to \Sigma_0^*$, where $\Sigma_0 = \Sigma - \Sigma_n$ ($\Sigma_n$ was introduced in Definition 2). Then, $P$ is GCC.

*Proof:* Assume $q_i = \delta(q_0, s), q_j = \delta(q_0, s')$ and $P(s) = P(s')$. Since $P$ is defined according to Definition 2, $q_j = \delta(q_i, \sigma_1^* \dots \sigma_k^*)$ is true, where $\{\sigma_1, \dots \sigma_k\} \subseteq \Sigma_n$. Assume $q_1 = \delta(q_i, \sigma_1^*), \dots, q_j = \delta(q_k, \sigma_k^*)$. From Definition 2, $q_i, q_m$, where $m = 1, \dots, k$, are ICC. Also, we know that $q_1 \in Q, q_2 \in Q_1, \dots, q_j \in Q_{k-1}$. Thus, $\forall k, q_k \in Q_{k-1}$ and $q_{k+1} \in Q_k$ are pairwise ICC. It means that, $q_i, q_j$ are also intrinsic control consistent. Therefore, $P$ is GCC.

$\square$

We prove that OCC and natural observer properties of $P$ for the closed and the marked languages of a plant implies that $P$ is GCC.

*Proposition 2:* Let $\mathbf{G} = (Q, \Sigma, \delta, q_0, Q_m)$ be a non-blocking plant, described by closed and marked languages $L, L_m \subseteq \Sigma^*$, respectively and $P: \Sigma^* \to \Sigma_0^*$ is defined. If $P$ is $L$-OCC and $L_m$- Observer, then $P$ is GCC.

*Proof:* Assume $q_i = \delta(q_0, s), q_j = \delta(q_0, s')$ and $P$ is $L$-OCC and $L_m$- Observer. We should prove that $(2-$

$i), (2 - ii)$ do hold. By definition of OCC property, assume $s = s'\sigma_1 ... \sigma_{k-1}, \sigma_i \in (\Sigma - \Sigma_0) \cap \Sigma_u$ $(i \leq k - 1)$ and there is $\sigma_k \in \Sigma_u \cap \Sigma_0$ such that $s\sigma_k$ is a string in $L$. Then, $P(s) = P(s'\sigma_1\sigma_2 ... \sigma_{k-1})$. We can write $P(s) = P(s')$. Since $\sigma_k$ is uncontrollable, $(2 - i)$ is automatically satisfied. Now, by definition of natural observer property, assume $s' \in L, s' \notin L_m$ and $s = s'u$. Then, $q_j \notin Q_m$. Thus, $(2 - ii)$ is automatically satisfied. The proof is complete.
□

*Corollary 1:* The reverse of Proposition 2 is not true, in general.

We are going to synthesize the optimal non-blocking decentralized supervisory control based on GCC property.

## 4. Synthesis the Optimal Non-blocking Decentralized Controller Using G-Control Consistency of a Natural Projection

Let **G** be a non-blocking plant, described by the closed and the marked languages $L, L_m \subseteq \Sigma^*$, respectively. Let the natural projection $P: \Sigma^* \rightarrow \Sigma_0^*$ is **G** − Control Consistent. A cover $\mathcal{G}$ can be constructed by Definition 3.

*Definition 3:* A cover $\mathcal{G} = \{Q_i \subseteq Q | i \in I, I \text{ is the index set}\}$ of $Q$ is a cover on **G**, if
$(i)\ (\forall i \in I)(\forall q, q' \in Q_i)(\exists s \neq s' \in \Sigma^*),$
$q = \delta(q_0, s), q' = \delta(q_0, s'), P(s) = P(s'), P \text{ is GCC},$
$(ii)\ (\forall q \in Q - \cup_i Q_i), q \in Q_i,$  (3)
$(iii)\ (\forall i \in I)(\forall \sigma \in \Sigma)(\exists j \in I), [(\forall q \in Q_i), \delta(q, \sigma)!$
$\Rightarrow \delta(q, \sigma) \in Q_j].$

A cover $\mathcal{G}$ lumps states of **G** into cells $Q_i$ ($i \in I$) if they have the same observation through the GCC projection channel. According to $(3 - ii)$, each cell of $\mathcal{G}$ is nonempty and each pair of states in one cell is ICC (because $P$ is GCC). According to $(3 - iii)$, all states that can be reached from any states in $Q_i$ by one step transition $\sigma$ is covered by some $Q_j$.

Now, a reduced plant $\mathbf{G_p} = (Q_p, \Sigma_0, \delta_p, q_{0,p}, Q_{m,p})$ can be constructed as follows,
$(i)\ q_{0,p} \in Q_p\ s.t.\ q_0 \in Q,$
$(ii)\ Q_{m,p} = \{i \in I | Q_i \cap Q_m \neq \emptyset\},$
$(iii)\ \delta_p: Q_p \times \Sigma_0 \rightarrow Q_p$ with $\delta_p(q_p, \sigma) = q_p'$,
provided for such choice of $q_p \in Q_j,$  (4)
$(\exists q_p \in Q_i)\ \delta_p(q_p, \sigma)$
$\in Q_i, (\forall q_p' \in Q_i)[\delta_p(q_p, \sigma) \in Q_j].$

Suppose that the specification is $E \subseteq \Sigma_0^*$, and $L_{m,p} := L_m(\mathbf{G_p}), L_p := L(\mathbf{G_p})$. Decentralized supervisor is calculated as follow,
$$K_0 := supC\left(E \cap P(L_{m,p}), P(L_p)\right). \quad (5)$$

Assume the recognizer of $K_0$ is $\mathbf{SUP_0} = (X_p, \Sigma_0, \xi_p, x_{0,p}, X_{m,p})$. We prove that $K_0$ is the optimal non-blocking decentralized controller for **G**.

*Lemma 1:* Let $\mathbf{G} = (Q, \Sigma, \delta, q_0, Q_m)$ be a non-blocking plant, and $\mathbf{G_p} = (Q_p, \Sigma_0, \delta_p, q_{0,p}, Q_{m,p})$ be constructed by (4). If $P: \Sigma^* \rightarrow \Sigma_0^*$ is GCC, and $\mathbf{SUP_0} = (X_p, \Sigma_0, \xi_p, x_{0,p}, X_{m,p})$ such that $K_0 = L_m(\mathbf{SUP_0})$, then, $P^{-1}(K_0) \cap L_m(\mathbf{G})$ is non-blocking.

*Proof:* Let $C_d := \mathbf{sync}(\mathbf{SUP_0}, \mathbf{G})$ be the recognizer of $P^{-1}(K_0) \cap L_m(\mathbf{G})$, such that $C_d = (X_d, \Sigma, \xi_d, x_{0,d}, X_{m,d})$. Assume $\exists s, s' \in \Sigma^*, x = \xi_d(x_{0,d}, s), x' = \xi_d(x_{0,d}, s')$ and $P(s) = P(s')$. Since $P$ is GCC, each pair of states $q, q' \in Q$ such that $q = \delta(q_0, s), q' = \delta(q_0, s')$, are ICC. Thus, $x$ and $x'$ are control consistent. It means that, if $\xi_d(x, \sigma)!$ then $\sigma$ cannot be disabled at $x'$, and vice versa. Therefore $P^{-1}(K_0) \cap L_m(\mathbf{G})$ is non-blocking.
□

In Theorem 1, we prove that $K_0$ is optimal (least restrictive). Namely, each string of events which is executed in $K$, can also be executed in $K_0 \| L_m$, where $L_m := L_m(\mathbf{G})$. Note that $K$ is calculated as follow,
$$K := supC(E \| L_m, L). \quad (6)$$

Also, assume $\mathbf{SUP} = (X, \Sigma, \xi, x_0, X_m)$ is the recognizer of $K$.

*Theorem 1:* Let **G** be a non-blocking plant, described by the closed and the marked languages $L := L(\mathbf{G}), L_m := L_m(\mathbf{G})$. Suppose that the specification is $E \subseteq \Sigma_0^*$. If the natural projection $P: \Sigma^* \rightarrow \Sigma_0^*$ is GCC, $K_0$ is calculated by (5) and $K$ is calculated by (6), then $K = K_0 \| L_m$.

*Proof:* We should prove that a. $K_0 \| L_m \subseteq K$, b. $K \subseteq K_0 \| L_m$.

Let $K_0$ and $K$ be calculated by (5) and (6), respectively. Since $K_0 = supC(E \cap P(L_{m,p}), P(L_p))$, we can write
$K_0 \| L_m \subseteq [E \cap P(L_{m,p})] \| L_m = P^{-1}[E \cap P(L_{m,p})] \cap L_m$
$= P^{-1}(E) \cap P^{-1}P(L_{m,p}) \cap L_m.$

From $(3 - i)$ some transitions in **G** are self-looped at some states in $\mathbf{G_p}$. Then, $L_m \subseteq L_{m,p}$. Thus, $P^{-1}(E) \cap P^{-1}P(L_{m,p}) \cap L_m = P^{-1}(E) \cap L_m = E \| L_m$.

Lemma 1 implies that $K_0 \| L_m$ is non-blocking. Thus, $\overline{K_0 \| L_m} = \overline{K_0} \| L$. Now, we prove that any string which occurs in $K_0 \| L_m$, also occurs in $K$. From (5), $K_0$ is controllable w.r.t. $P(L_p)$. Also, $P(L_p) \| L = P^{-1}P(L_p) \cap L = L$, Because $L \subseteq L_p$. Thus, we can write $K_0 \| L_m$ is controllable w.r.t. $L$. It means that
$K_0 \| L_m = supC\left(E \cap P(L_{m,p}), P(L_p)\right) \| L_m$
$\subseteq supC(E \| L_m, L) = K$

Now, we prove $K \subseteq K_0 \| L_m$.
(b) $K \subseteq P^{-1}(K_0) \cap L_m(\mathbf{G})$

We prove that any string, which is not in $P^{-1}(K_0) \cap L_m(\mathbf{G})$, is not in $K$. Since $P$ is GCC, there exists $\exists s, s' \in \Sigma^*$ such that $P(s) = P(s')$ and $\delta(q_0, s), \delta(q_0, s') \in Q$. If $\delta(q_0, s) \in Q_m$, then $\delta(q_0, s') \notin Q_m$ and vice versa. Thus, we can write $\xi(x_0, s') \notin X_m$. Since $s' \notin L_m(\mathbf{G})$, we can write $s' \notin P^{-1}(K_0) \cap L_m(\mathbf{G})$.
□

Theorem 1 can be verified by synthesis a decentralized surge-avoidance supervisor for a gas compressor station.

## 5. Synthesis the Optimal Non-blocking Decentralized Supervisory Control for Surge-Avoidance in a Gas Compressor Station

Surge is a symmetric oscillation of gas flow through a compressor, when the gas flow is lower than the minimum flow. Surge can also induce vibrations in other components of the compression system, such as connected piping. The surge control system is an important element in the compressor system. Protection of the compressor through the surge control system helps to avoid repetitive repairs and maintenance works. However, the primary objective of any surge control system should be to predict and to prevent the occurrence of surge to reduce possible damage to the compressor and ensure a safe working environment for all station personnel, the principle of a centrifugal compressor surge control system is based on ensuring that the gas flow through the compressor is not reduced below the minimum flow at a specific head. The minimum flow is provided by opening the recycle valve, which is parallel to each compressor. The recycle valve should be opened in a pre-specified time by the control system. The command signal to the valve is calculated based on the compressor operation, its proximity and its movement relevant to the surge control line (Fig. 1). Various design approaches are proposed for surge control system (Gravdahl & Egeland, 1999). In (Saeidi, Afzalian & Gharavian, submitted to IEEE/ASME Trans. Mechatronics), the "design to avoid surge" approach was considered.

Consider a DES model for a plant with a recycle valve including PI controller and a compressor, under surge condition (Fig. 2). In order to save energy in the compressor station, it is preferable to close the recycle valve if the compressor is in the normal operation. The control logic for surge-avoidance in a compressor station is shown in Fig. 3 (Saeidi, Afzalian & Gharavian, submitted to IEEE/ASME Trans. Mechatronics). The plant and the specification models are implemented in TCT software (Wonham, 2014). Suppose that $\Sigma = \{12,14,51,52,53,54,55,57\}$, and the controllable events set is $\Sigma_c = \{51,53,55,57\}$. The supervisor $K = L_m(\textbf{SUP})$ is synthesized for surge-avoidance in a gas compressor station (Fig. 4).

Moreover, $P: \Sigma^* \rightarrow \Sigma_0^*$ is GCC, where $\Sigma_0 = \Sigma - \{52,54\}$. The projection of the reduced plant, $\textbf{PG}_\textbf{p}$ is shown in Fig. 5.

Having $P$, we can synthesize decentralized supervisor $\textbf{SUP}_\textbf{0}$, using **supcon** procedure in TCT on the projection of the plant, $\textbf{PG}_\textbf{p}$ (Fig. 6). Obviously, $\textbf{sync}(\textbf{SUP}_\textbf{0}, \textbf{G}) = \textbf{SUP}$.

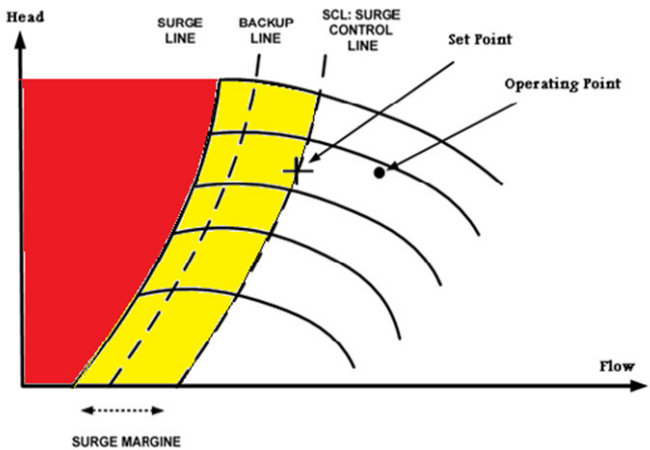

Fig. 1. Surge map for a surge-avoidance control system

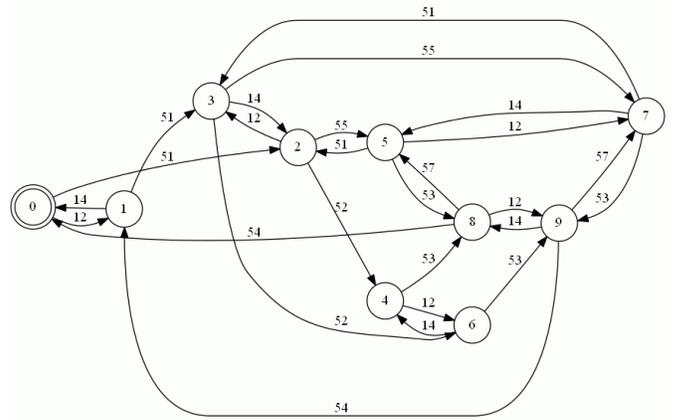

Fig. 2. DES model of the Plant including recycle valve, PI controller and a compressor (**G**)

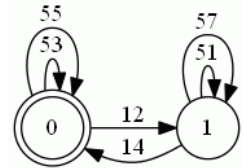

Fig. 3. The specification for surge-avoidance (**E**)

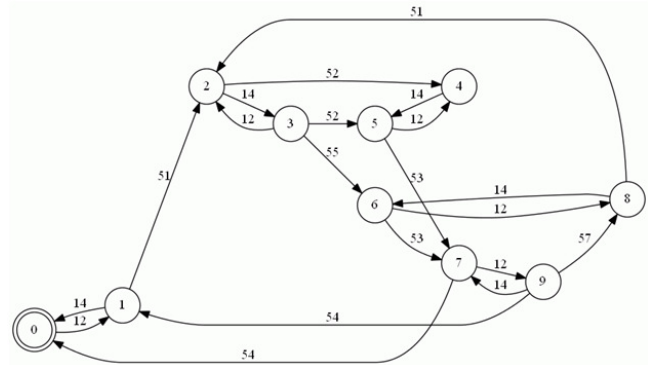

Fig. 4. The monolithic supervisor, **SUP**

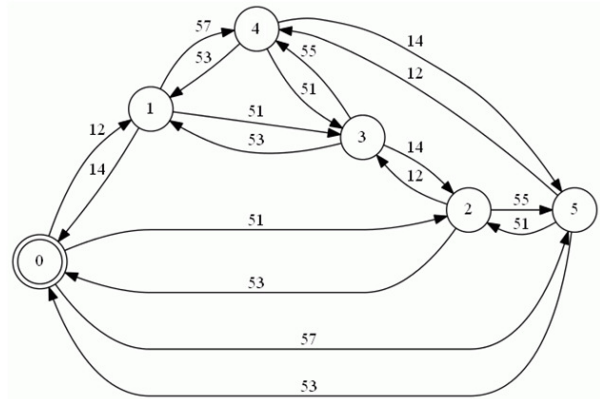

Fig. 5. Natural projection of the reduced plant ($\textbf{PG}_\textbf{p}$)

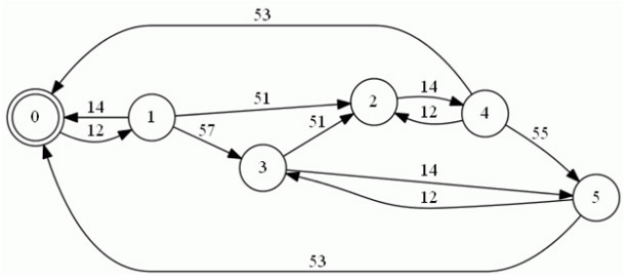

Fig. 6. Decentralized supervisory controller, $SUP_0$

## 6. Conclusions

This paper addresses construction of the optimal non-blocking decentralized supervisory control based on GCC property for a natural projection w.r.t. the plant. For this purpose, we defined ICC property for each pair of states in the plant. Also, we defined GCC property for a natural projection. We proved that if a natural projection is OCC for the closed language of the plant and is natural observer for the marked language of the plant, then it is GCC. Thus, we relaxed conditions to synthesize the optimal non-blocking decentralized supervisory control of DES. We proved that, if a natural projection is GCC, then the supremal controllable sublanguage of the projection of the reduced plant is equivalent to the supremal controllable sublanguage of synchronization of the specification with the plant. We employed the proposed technique for synthesis the optimal non-blocking decentralized supervisory control for surge-avoidance in a typical gas compressor station.